\documentclass[aps,prl,reprint,10pt,fleqn,showpacs,showkeys,floatfix,superscriptaddress]{revtex4-1}
\usepackage{graphicx}
\usepackage{amssymb}
\usepackage{amsmath}
\usepackage{bm}
\usepackage{dcolumn}

\begin{document}


\title{Unusual Magnetic Response of an $S = 1$ Antiferromagetic Linear-Chain Material}

\author{Jian-Sheng Xia}
\affiliation{Department of Physics and the National High Magnetic Field Laboratory, University of Florida, 
Gainesville, FL 32611-8440, USA}
\author{Andrzej Ozarowski}
\affiliation{National High Magnetic Field Laboratory, Florida State University, Tallahassee, FL 32310, USA }
\author{Peter M.~Spurgeon}
\affiliation{Department of Chemistry and Biochemistry, Eastern Washington University, Cheney, WA 99004, USA}
\author{Adora~G.~Baldwin}
\affiliation{Department of Chemistry and Biochemistry, Eastern Washington University, Cheney, WA 99004, USA}
\author{Jamie L.~Manson}
\affiliation{Department of Chemistry and Biochemistry, Eastern Washington University, Cheney, WA 99004, USA}
\author{Mark W.~Meisel}
\affiliation{Department of Physics and the National High Magnetic Field Laboratory, University of Florida, 
Gainesville, FL 32611-8440, USA}

\date{\today}

\begin{abstract}
\vskip 0.4cm
An $S=1$ antiferromagnetic polymeric chain, [Ni(HF$_2$)(3-Clpy)$_4$]BF$_4$ (py = pyridine), has previously been 
identified to have intrachain, nearest-neighbor antiferromagnetic interaction strength 
$J/k_{\mathrm{B}} = 4.86$~K and single-ion anisotropy (zero-field splitting) $D/k_{\mathrm{B}} = 4.3$~K, so the 
ratio $D/J = 0.88$  places this system close to the $D/J \approx 1$ gapless critical point between the  
topologically distinct Haldane and Large-$D$ phases.  The magnetization was studied over a range of temperatures, 
50~mK~$\leq T \leq 1$~K, and magnetic fields, $B \leq 10$~T, in an attempt to identify a critical field, 
$B_{\mathrm{c}}$, associated with the closing of the Haldane gap, and the present work places an upper bound of 
$B_{\mathrm{c}} \leq (35 \pm 10)$~mT.  At higher fields, the observed magnetic response is qualitatively 
similar to the ``excess'' signal observed by other workers at 0.5~K and below 3~T.  The high-field 
(up to 14.5~T), multi-frequency (nomially 200~GHz to 425~GHz) ESR spectra at 3~K reveal several broad features 
considered to be associated with the linear-chain sample.
 
\end{abstract}

\maketitle


After Haldane reported the fundamental differences between one-dimensional Heisenberg antiferromagnets with 
integer and half-integer spins \cite{Haldane1,Haldane2}, theoretical \cite{Schulz,Affleck1,Affleck2} and 
experimental \cite{Yamashita} activities focused on clarifying the magnetic phases and properties of a 
broad range of linear-chain magnets described by the Hamiltonian
\begin{equation}
\label{Hamiltonian}
\begin{split}
\mathcal{H}\;=\;J\,\sum_i \{[S^x_i S^x_{i+1}\;+\,S^y_i S^y_{i+1}\;+\;\lambda\,S^z_i S^z_{i+1}] \\
+ \;D\,(S^z_1)^2\;+\;E\, [(S^x_i)^2-(S^y_i)^2]\} \;-\; \vec{B} \,\cdot 
\stackrel{\leftrightarrow}{g} \cdot\, \vec{S}   \;\;\;\;,
\end{split}
\end{equation}
where $J$ is the nearest-neighbor intrachain magnetic exchange parameter, $D$ and $E$ are the single-ion and 
rhombic anisotropies that are commonly referred to as zero-field splitting parameters arising from the 
crystal-field anisotropy, $g$ 
is the Lande $g$-factor tensor, $B$ is the applied magnetic field, $\lambda$ is a parameter distorting the 
exchange, and the inevitable interchain coupling $J^{\prime}$ is not shown. 
For $S=1$, the Haldane phase, which possesses an energy gap $\Delta = 0.41 J$ when $D=0$, and the Large-$D$ 
phase \cite{Schulz,Botet} have been established by magnetization \cite{Renard,Orendac1}, 
EPR \cite{Orendac2,Sieling,Hagiwara1,Cizmar}, and inelastic neutron scattering studies \cite{Renard,Zheludev}.  
Although several materials have been identified in both limits, to date, no real material has been identified 
as being close to the quantum critical point of $D/J \approx 1$.  This region of the phase diagram has attracted 
wide theoretical and numerical attention 
\cite{Sakai,Tzeng,Albuquerque,Pollmann1,Hu,Pollmann2,Li,Langari,Wierschem} and the ground-state properties have 
not been unambigously resolved.  Nonetheless, there is converging consensus that the Haldane gap closes 
and the Large-$D$ gap opens at a gapless critical point, $(D/J)_{\mathrm{c}} = 0.97$ when $\lambda = 1$, between the 
two topologically distinct gapped phases with different parity.  In other words, this particular quantum critical 
region, which has attracted intense theoretical and numerical scrutiny, remains without any candidate systems 
whose properties may provide insight into the measurable thermodynamic and electromagnetic properties.  Resolving 
this lacuna will provide new perspectives into quantum critical phenomena, and such experimental-theoretical advances 
are being realized in other systems \cite{Steglich,Kinross}.       

\begin{figure}[t]
\includegraphics[width=7.5cm]{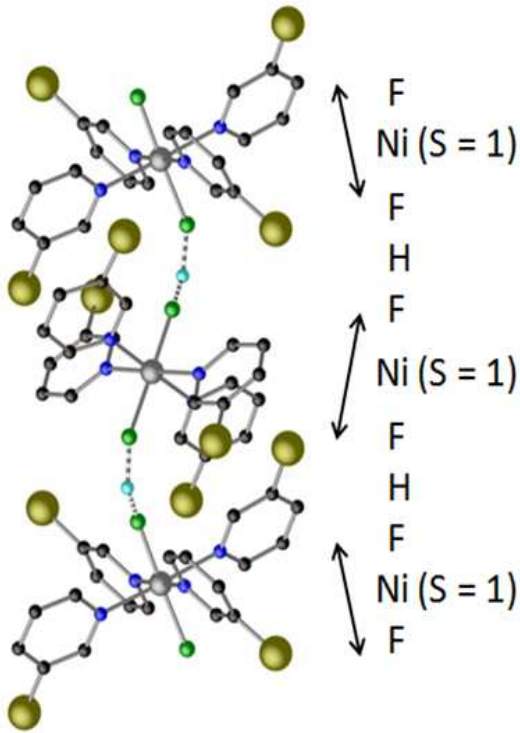}
\caption{\label{Crystal}(Color online) The crystal structure of [Ni(HF$_2$)(3-Clpy)$_4$]BF$_4$ (py = pyridine), 
space group $P2_1/c$, resembles a linear chain of octahedrally coordinated Ni$^{2+}$ ions linked by F$-$H$-$F 
bridges along the $c$-axis \cite{Manson}. The 3-Clpyridine (C$_5$H$_4$ClN) ligands isolate the chains from 
neighboring ones, and the BF$_4$ anions required for charge neutralization are omitted for clarity. The local environment 
of the Ni ions switches axial orientation along the chain, as sketched by the arrows, and is an 
aspect reminescint of the staggered $g$-tensor in Ni(C$_2$H$_8$N$_2$)$_2$NO$_2$(ClO$_4$) (commonly referred to 
as NENP) \cite{Chiba,Affleck-staggered}.}
\end{figure}

The purpose of this paper is to report new experimental magnetization and EPR data on an intriguing 
$S=1$ polymeric chain material, [Ni(HF$_2$)(3-Clpy)$_4$]BF$_4$ (py = pyridine) \cite{Manson}, Fig.~\ref{Crystal}, 
that appears to be located close to $(D/J)_{\mathrm{c}}$.  In 2012, 
[Ni(HF$_2$)(3-Clpy)$_4$]BF$_4$ was synthesized and characterized by a wide-range of spectroscopies \cite{Manson}. 
Most notably, muon-spin relaxation spectra relax monotonically, and the lack of oscillations or a shift in the 
base line are interpreted as indicating the absence of long-range magnetic order.  The low-field $B = 0.1$~T, 
high-temperature $(T \geq 1.8$~K) magnetic susceptibility possessed an antiferromagnetic component, which could be 
fit to $J/k_{\mathrm{B}} = 4.86$~K with $g=2.1$, and a paramagnetic Curie-like component that might be associated 
with uncoupled Ni$^{2+}$ spins and/or with $S=1/2$ end-chain spins 
\cite{Affleck2,Hagiwara2,Glarum,Avenel1,Avenel2,Batista,Granroth}.  In addition, analysis of the UV-vis spectra suggested 
$D/k_{\mathrm{B}} \approx 4.3$~K, while heat capacity data are consistent with $D/J \lesssim 1$.  
Finally, magnetization data at 
$T = 0.5$~K and in $B < 3$~T revealed the presence of an ``excess'' amount of signal above the response measured 
at 1.6~K.  The present studies were initiated to search for evidence of the critical field required to close 
the Haldane gap and to understand the origin of the magnetic signal at low temperatures.

\begin{figure}[t]
\includegraphics[width=5.3cm]{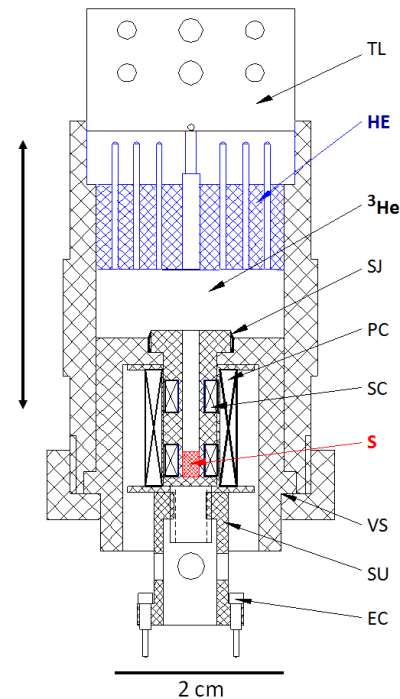}
\caption{\label{Exp-Cell}(Color online) The experimental cell used for magnetization studies is mostly made of 
plastics and is oriented along the direction of the magnetic fields represented by the left arrow.  The main 
components are: a Ag thermal link (TL), with a Ag sintered powder heat exchanger (HE), a bath of pure $^3$He 
(fill line not shown) in direct contact with the sample (S), a Stycast 1266 joint (SJ) to the Kel-F 
(Neoflon PCTFE) housing, a primary coil (PC) consisting of 2100 turns of nominally 76~$\mu$m diameter NbTi 
multi-filamentary wire, a set of secondary coils (SC) each with 2500 turns of 51~$\mu$m diameter Cu wire, 
a lower plastic-to-plastic (reusable) vacuum seal (VS), and a supporting tower (SU) for the electrical 
connections (EC).}
\end{figure}

To experimentally explore the magnetization of [Ni(HF$_2$)(3-Clpy)$_4$]BF$_4$, nominally $4~(\pm 1)$~mg of small 
cubic-like (with sides of a few mm), as-grown \cite{Manson} polycrystalline pieces were randomly placed and rigidly 
held in one side of a mutual inductance cell, shown in Fig.~\ref{Exp-Cell}, mounted to the dilution refrigerator 
operated in the Williamson Hall Annex of the High-B/T Facility of the National High Magnetic Field Laboratory 
(NHMFL) at the University of Florida.  At a fixed temperature, the imbalance between the two secondary coils was 
monitored by a lock-in amplifier (Model NF5610B) at low frequency (52.93~Hz) and low ac field (0.1~mT) while 
sweeping a superconducting solenoid at a constant rate of 50~mT/min.  The phase setting was adjusted to minimize 
one channel of the lock-in amplifier and remained constant for all runs.  The data acquisition of the in-phase 
and out-of-phase voltages 
($V_{\mathrm{in}}$ and $V_{\mathrm{out}}$) takes about 3.5 hours as the field is swept to or from 10~T.  To 
within experimental resolution, no magnetic hysteresis was established at 50~mK, but the starting and final 
voltages differ by a few $\mu$V due to changes in the level of the liquid helium in the bath.  With the 
absence of detectable hystersis at the lowest temperature, the runs at higher temperatures were performed 
by sweeping the field in only one direction.  When the runs with the sample were complete, the sample was 
removed, and the measurements were repeated so ``sample-in minus sample-out'' analysis, commonly employed 
in pulsed field experiments, could be preformed. It is noteworthy the sample was cooled by a liquid of pure $^3$He, 
which is known to provide excellent thermal coupling down to low temperatures and in high magnetic fields 
\cite{Hammel,Schuhl}, and a similar  arrangement was used to cool another molecule-based magnetic system down 
to 40~mK \cite{Orendacova}.  In fact, the solvent associated with the present sample \cite{Manson} eroded 
the thin Stycast 1266 wall of the previous cell, and a new one was constructed from an old piece of Kel-F by 3M 
(now available as Daikin Neoflon PCTFE).

\begin{figure}[b]
\includegraphics[width=8.6cm]{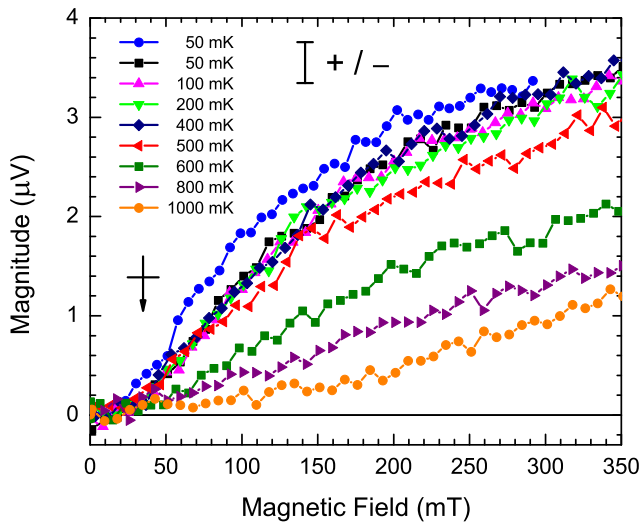}
\caption{\label{Magn-350mT}(Color online) The magnitudes of the measured voltages, after background subtraction, 
are shown as a function of the magnetic field, and the overall uncertainty is represented by the bar designating 
``$+/-$''.  The legend identifies the temperatures, and two runs at 50~mK are discussed in the text.  The 
apparent degeneracy of the data sets for $T \lesssim 400$~mK is conjectured to be evidence of the increase of 
entropy near the critical point, as discussed in the text.  The critical field necessary to close the Haldane 
gap has an upper bound of $B_{\mathrm{c}} \leq (35 \pm 10)$~mT, as indicated by the arrow and crossbar.}
\end{figure}

The isothermal, low-field dependences of the magnitudes $(= \sqrt{V_{\mathrm{in}}^2+V_{\mathrm{out}}^2})$, 
after background subtraction, are shown in Fig.~\ref{Magn-350mT}.  The overall experimental uncertainty, which 
is a combination of background subtraction and the bath level adjustment, is represented by the bar designating 
``$+/-$''.  The legend identifies the temperatures, and two runs at 50~mK were made.  One 50~mK run was made at 
the start of the experiment (shown in blue), while the other 50~mK run (shown in black), which followed the 
measurement at 1000~mK, was made at the end of the study with the sample present.  One striking aspect of the 
data is the apparent similarity of the data sets for $T \lesssim 400$~mK, and at first glance, this result 
seems to indicate the sample is not being cooled to the lowest temperatures.  However, it is important to 
stress several points.  Firstly, as previously mentioned, similar samples have been reproducibly cooled in 
other studies \cite{Orendacova}.  Secondly, difficulty in cooling a sample may indicate an increase of the 
specific heat near a quantum phase transition, where the accumulation of entropy is generally accepted as a 
generic consequence \cite{Vidal,Si2}. In addition, the data at the lowest fields suggest the magnetic signal 
is absent or exceedingly small.  Although the sensitivity of this cell is considered to be comparable to the 
one used in other work \cite{Orendacova}, the dynamic range of the detection coils has not been established at 
this time, so a definitive value for the critical field necessary to close the Haldane gap, $B_{\mathrm{c}}$, 
cannot be established.  Nevertheless, an upper bound of $B_{\mathrm{c}} \leq (35 \pm 10)$~mT can be set as shown 
in Fig.~\ref{Magn-350mT}.  This result can be compared to the extrapolated Haldane gap values near 
$(D/J)_{\mathrm{c}} = 0.968$, as given by Hu \emph{et al.}~\cite{Hu}, and the result is $B_{\mathrm{c}} = 46$~mT, 
when using the aforementioned values for the parameters.  Finally, the ``excess'' signal, at $T = 0.5$~K and 
$B < 3$~T, reported by others \cite{Manson} has been qualitatively reproduced in this work, but quantitative 
agreement between the two studies is lacking.  Furthermore, the temperature-field dependences of the data sets 
shown in Fig.~\ref{Magn-350mT} cannot be modeled by non-interacting $S=1/2$ end-chain spins.  Consequently, the 
magnetic response at low temperatures and in low magnetic fields is conjectured to be associated with the 
ground state susceptibility, but quantitative analysis will require data acquired with oriented crystals, 
where the impact of a staggered $g$-tensor, Fig.~\ref{Crystal}, may be revealed \cite{Chiba,Affleck-staggered}.

\begin{figure}[t]
\includegraphics[width=8.6cm]{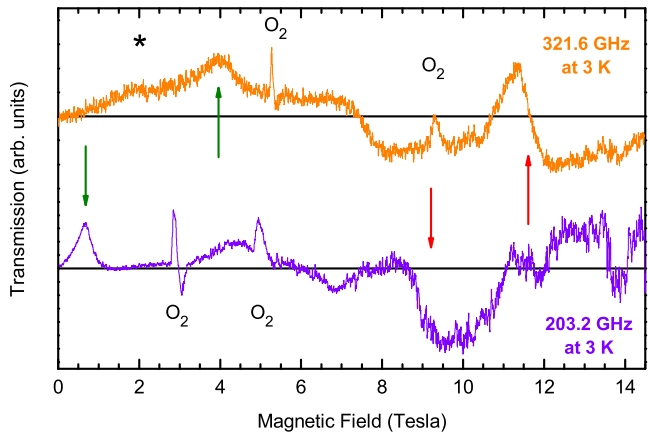}
\caption{\label{EPR}(Color online) The EPR transmission signals at 321.6~GHz (top and organge trace) and 
203.2~GHz (bottom and purple trace) are shown as a function of magnetic field.  For each frequency, the 
arrows designate the features considered to arise from the Ni-chain, while fingerprints 
known to be indicative of trace amounts of oxygen in solid air are labeled as O$_2$ \cite{Pardi}.  A weak 
feature, which remains unidentified, is marked by an asterisk.}
\end{figure}

High-field, high-frequency EPR spectra were acquired with a randomly-packed, purposely-powdered sample, with 
a total mass of approximately 50~mg, that was synthesized separately from the ones used in the magnetization 
studies.  Using a constant flow cryostat operating at 3~K $\leq T \leq 20$~K, the experiments were conducted 
with a home-built spectrometer at the electromagnetic resonance (EMR) facility of the NHMFL at Florida State University.  
The details of the transmission-type instrument and spectrometer are described elsewhere 
\cite{Hassan}, and for this work, spectra were acquired at several frequencies ranging from 203~GHz to 425~GHz.  
The spectra taken at $T=3$~K and at 203.2~GHz and 321.6~GHz are shown in Fig.~\ref{EPR}, where the broad features 
associated with the Ni-chains are identified.
Since the thermal energy at $T=3$~K is similar to the $J$ and $D$ energies, the full dynamics of the Haldane 
chain are not developed, especially when considering the Haldane gap is exceedingly small, 
$\Delta \lesssim 60$~mK, if it exists.  The limited EPR data sets preclude any extended analysis with various 
reasonable or far-fetched assumptions, e.g. short-range or long-range antiferromagnetic order is 
present \cite{Sieling,Hagiwara1,Cizmar,HagiwaraMnF2,Fanucci,PhysRevB.83.224417}.  Ultimately, future EPR 
investigations will need 
to employ oriented single crystal samples that can be studied over a wider range of frequency and temperature.

In summary, the magnetic properties of [Ni(HF$_2$)(3-Clpy)$_4$]BF$_4$ (py = pyridine) exhibit unusual behavior 
at low temperatures and in low fields.  An upper bound for the critical field required to 
close the Haldane gap is established, $B_{\mathrm{c}} \leq (35 \pm 10)$~mT, and this value is close to the 
predicted one \cite{Hu}, 46~mT, when using the experimentally established values of $J$, $D$, and $g$.  
In low fields, the magnetic signal increases with decreasing temperature for 400~mK $\lesssim T \lesssim 800$~mK but, 
to within experimental resolution, is independent of temperature for 50~mK $\leq T \leq 400$~mK.  
This observation is consistent with a significant increase in the specific heat arising 
from the accumulation of entropy in the vicinity of the quantum critical transition near $D/J \approx 1$.  
These results provide new experimental access to this quantum critical regime, where the $D/J$ value might be tunable 
by pressure \cite{Broholm,Oosawa,Ruegg,Tao,Zheludev2} and/or chemical modifications such as deuteration 
\cite{Takano,Goddard}.  Ultimately, in the future, a full suite of experimental tools on oriented single 
crystal samples will provide insight into this magnetolomic cornucopia.    

\begin{acknowledgments}
This work was supported, in part, by the National Science Foundation through DMR-1202033 (MWM), DMR-1306158 
(JLM), and DMR-1157490 (NHMFL).  We acknowledge enlightening communications with M.~Hagiwara, M.~Orend\'{a}\v{c}, 
A.~Orend\'{a}\v{c}ov\'{a}, and E.~\v{C}i\v{z}m\'{a}r.
\end{acknowledgments}
\bibliography{MansonNiv20sept}

\begin{thebibliography}{51}%
\makeatletter
\providecommand \@ifxundefined [1]{%
 \@ifx{#1\undefined}
}%
\providecommand \@ifnum [1]{%
 \ifnum #1\expandafter \@firstoftwo
 \else \expandafter \@secondoftwo
 \fi
}%
\providecommand \@ifx [1]{%
 \ifx #1\expandafter \@firstoftwo
 \else \expandafter \@secondoftwo
 \fi
}%
\providecommand \natexlab [1]{#1}%
\providecommand \enquote  [1]{``#1''}%
\providecommand \bibnamefont  [1]{#1}%
\providecommand \bibfnamefont [1]{#1}%
\providecommand \citenamefont [1]{#1}%
\providecommand \href@noop [0]{\@secondoftwo}%
\providecommand \href [0]{\begingroup \@sanitize@url \@href}%
\providecommand \@href[1]{\@@startlink{#1}\@@href}%
\providecommand \@@href[1]{\endgroup#1\@@endlink}%
\providecommand \@sanitize@url [0]{\catcode `\\12\catcode `\$12\catcode
  `\&12\catcode `\#12\catcode `\^12\catcode `\_12\catcode `\%12\relax}%
\providecommand \@@startlink[1]{}%
\providecommand \@@endlink[0]{}%
\providecommand \url  [0]{\begingroup\@sanitize@url \@url }%
\providecommand \@url [1]{\endgroup\@href {#1}{\urlprefix }}%
\providecommand \urlprefix  [0]{URL }%
\providecommand \Eprint [0]{\href }%
\providecommand \doibase [0]{http://dx.doi.org/}%
\providecommand \selectlanguage [0]{\@gobble}%
\providecommand \bibinfo  [0]{\@secondoftwo}%
\providecommand \bibfield  [0]{\@secondoftwo}%
\providecommand \translation [1]{[#1]}%
\providecommand \BibitemOpen [0]{}%
\providecommand \bibitemStop [0]{}%
\providecommand \bibitemNoStop [0]{.\EOS\space}%
\providecommand \EOS [0]{\spacefactor3000\relax}%
\providecommand \BibitemShut  [1]{\csname bibitem#1\endcsname}%
\let\auto@bib@innerbib\@empty
\bibitem [{\citenamefont {Haldane}(1983{\natexlab{a}})}]{Haldane1}%
  \BibitemOpen
  \bibfield  {author} {\bibinfo {author} {\bibfnamefont {F.}~\bibnamefont
  {Haldane}},\ }\href {\doibase http://dx.doi.org/10.1016/0375-9601(83)90631-X}
  {\bibfield  {journal} {\bibinfo  {journal} {Physics Letters A}\ }\textbf
  {\bibinfo {volume} {93}},\ \bibinfo {pages} {464 } (\bibinfo {year}
  {1983}{\natexlab{a}})}\BibitemShut {NoStop}%
\bibitem [{\citenamefont {Haldane}(1983{\natexlab{b}})}]{Haldane2}%
  \BibitemOpen
  \bibfield  {author} {\bibinfo {author} {\bibfnamefont {F.~D.~M.}\
  \bibnamefont {Haldane}},\ }\href {\doibase 10.1103/PhysRevLett.50.1153}
  {\bibfield  {journal} {\bibinfo  {journal} {Phys. Rev. Lett.}\ }\textbf
  {\bibinfo {volume} {50}},\ \bibinfo {pages} {1153} (\bibinfo {year}
  {1983}{\natexlab{b}})}\BibitemShut {NoStop}%
\bibitem [{\citenamefont {Schulz}(1986)}]{Schulz}%
  \BibitemOpen
  \bibfield  {author} {\bibinfo {author} {\bibfnamefont {H.~J.}\ \bibnamefont
  {Schulz}},\ }\href {\doibase 10.1103/PhysRevB.34.6372} {\bibfield  {journal}
  {\bibinfo  {journal} {Phys. Rev. B}\ }\textbf {\bibinfo {volume} {34}},\
  \bibinfo {pages} {6372} (\bibinfo {year} {1986})}\BibitemShut {NoStop}%
\bibitem [{\citenamefont {Affleck}\ \emph {et~al.}(1987)\citenamefont
  {Affleck}, \citenamefont {Kennedy}, \citenamefont {Lieb},\ and\ \citenamefont
  {Tasaki}}]{Affleck1}%
  \BibitemOpen
  \bibfield  {author} {\bibinfo {author} {\bibfnamefont {I.}~\bibnamefont
  {Affleck}}, \bibinfo {author} {\bibfnamefont {T.}~\bibnamefont {Kennedy}},
  \bibinfo {author} {\bibfnamefont {E.~H.}\ \bibnamefont {Lieb}}, \ and\
  \bibinfo {author} {\bibfnamefont {H.}~\bibnamefont {Tasaki}},\ }\href
  {\doibase 10.1103/PhysRevLett.59.799} {\bibfield  {journal} {\bibinfo
  {journal} {Phys. Rev. Lett.}\ }\textbf {\bibinfo {volume} {59}},\ \bibinfo
  {pages} {799} (\bibinfo {year} {1987})}\BibitemShut {NoStop}%
\bibitem [{\citenamefont {Affleck}(1989)}]{Affleck2}%
  \BibitemOpen
  \bibfield  {author} {\bibinfo {author} {\bibfnamefont {I.}~\bibnamefont
  {Affleck}},\ }\href {http://stacks.iop.org/0953-8984/1/i=19/a=001} {\bibfield
   {journal} {\bibinfo  {journal} {Journal of Physics: Condensed Matter}\
  }\textbf {\bibinfo {volume} {1}},\ \bibinfo {pages} {3047} (\bibinfo {year}
  {1989})}\BibitemShut {NoStop}%
\bibitem [{\citenamefont {Yamashita}\ \emph {et~al.}(2000)\citenamefont
  {Yamashita}, \citenamefont {Ishii},\ and\ \citenamefont
  {Matsuzaka}}]{Yamashita}%
  \BibitemOpen
  \bibfield  {author} {\bibinfo {author} {\bibfnamefont {M.}~\bibnamefont
  {Yamashita}}, \bibinfo {author} {\bibfnamefont {T.}~\bibnamefont {Ishii}}, \
  and\ \bibinfo {author} {\bibfnamefont {H.}~\bibnamefont {Matsuzaka}},\ }\href
  {\doibase http://dx.doi.org/10.1016/S0010-8545(99)00212-X} {\bibfield
  {journal} {\bibinfo  {journal} {Coordination Chemistry Reviews}\ }\textbf
  {\bibinfo {volume} {198}},\ \bibinfo {pages} {347 } (\bibinfo {year}
  {2000})}\BibitemShut {NoStop}%
\bibitem [{\citenamefont {Botet}\ \emph {et~al.}(1983)\citenamefont {Botet},
  \citenamefont {Jullien},\ and\ \citenamefont {Kolb}}]{Botet}%
  \BibitemOpen
  \bibfield  {author} {\bibinfo {author} {\bibfnamefont {R.}~\bibnamefont
  {Botet}}, \bibinfo {author} {\bibfnamefont {R.}~\bibnamefont {Jullien}}, \
  and\ \bibinfo {author} {\bibfnamefont {M.}~\bibnamefont {Kolb}},\ }\href
  {\doibase 10.1103/PhysRevB.28.3914} {\bibfield  {journal} {\bibinfo
  {journal} {Phys. Rev. B}\ }\textbf {\bibinfo {volume} {28}},\ \bibinfo
  {pages} {3914} (\bibinfo {year} {1983})}\BibitemShut {NoStop}%
\bibitem [{\citenamefont {Renard}\ \emph {et~al.}(1987)\citenamefont {Renard},
  \citenamefont {Verdaguer}, \citenamefont {Regnault}, \citenamefont
  {Erkelens}, \citenamefont {Rossat-Mignod},\ and\ \citenamefont
  {Stirling}}]{Renard}%
  \BibitemOpen
  \bibfield  {author} {\bibinfo {author} {\bibfnamefont {J.~P.}\ \bibnamefont
  {Renard}}, \bibinfo {author} {\bibfnamefont {M.}~\bibnamefont {Verdaguer}},
  \bibinfo {author} {\bibfnamefont {L.~P.}\ \bibnamefont {Regnault}}, \bibinfo
  {author} {\bibfnamefont {W.~A.~C.}\ \bibnamefont {Erkelens}}, \bibinfo
  {author} {\bibfnamefont {J.}~\bibnamefont {Rossat-Mignod}}, \ and\ \bibinfo
  {author} {\bibfnamefont {W.~G.}\ \bibnamefont {Stirling}},\ }\href
  {http://stacks.iop.org/0295-5075/3/i=8/a=013} {\bibfield  {journal} {\bibinfo
   {journal} {EPL (Europhysics Letters)}\ }\textbf {\bibinfo {volume} {3}},\
  \bibinfo {pages} {945} (\bibinfo {year} {1987})}\BibitemShut {NoStop}%
\bibitem [{\citenamefont {Orend\'a\v{c}}\ \emph {et~al.}(1995)\citenamefont
  {Orend\'a\v{c}}, \citenamefont {Orend\'a\v{c}ov\'a}, \citenamefont
  {\v{C}ern\'ak}, \citenamefont {Feher}, \citenamefont {Signore}, \citenamefont
  {Meisel}, \citenamefont {Merah},\ and\ \citenamefont {Verdaguer}}]{Orendac1}%
  \BibitemOpen
  \bibfield  {author} {\bibinfo {author} {\bibfnamefont {M.}~\bibnamefont
  {Orend\'a\v{c}}}, \bibinfo {author} {\bibfnamefont {A.}~\bibnamefont
  {Orend\'a\v{c}ov\'a}}, \bibinfo {author} {\bibfnamefont {J.}~\bibnamefont
  {\v{C}ern\'ak}}, \bibinfo {author} {\bibfnamefont {A.}~\bibnamefont {Feher}},
  \bibinfo {author} {\bibfnamefont {P.~J.~C.}\ \bibnamefont {Signore}},
  \bibinfo {author} {\bibfnamefont {M.~W.}\ \bibnamefont {Meisel}}, \bibinfo
  {author} {\bibfnamefont {S.}~\bibnamefont {Merah}}, \ and\ \bibinfo {author}
  {\bibfnamefont {M.}~\bibnamefont {Verdaguer}},\ }\href {\doibase
  10.1103/PhysRevB.52.3435} {\bibfield  {journal} {\bibinfo  {journal} {Phys.
  Rev. B}\ }\textbf {\bibinfo {volume} {52}},\ \bibinfo {pages} {3435}
  (\bibinfo {year} {1995})}\BibitemShut {NoStop}%
\bibitem [{\citenamefont {Orend\'a\v{c}}\ \emph {et~al.}(1999)\citenamefont
  {Orend\'a\v{c}}, \citenamefont {Zvyagin}, \citenamefont {Orend\'a\v{c}ov\'a},
  \citenamefont {Sieling}, \citenamefont {L\"uthi}, \citenamefont {Feher},\
  and\ \citenamefont {Meisel}}]{Orendac2}%
  \BibitemOpen
  \bibfield  {author} {\bibinfo {author} {\bibfnamefont {M.}~\bibnamefont
  {Orend\'a\v{c}}}, \bibinfo {author} {\bibfnamefont {S.}~\bibnamefont
  {Zvyagin}}, \bibinfo {author} {\bibfnamefont {A.}~\bibnamefont
  {Orend\'a\v{c}ov\'a}}, \bibinfo {author} {\bibfnamefont {M.}~\bibnamefont
  {Sieling}}, \bibinfo {author} {\bibfnamefont {B.}~\bibnamefont {L\"uthi}},
  \bibinfo {author} {\bibfnamefont {A.}~\bibnamefont {Feher}}, \ and\ \bibinfo
  {author} {\bibfnamefont {M.~W.}\ \bibnamefont {Meisel}},\ }\href {\doibase
  10.1103/PhysRevB.60.4170} {\bibfield  {journal} {\bibinfo  {journal} {Phys.
  Rev. B}\ }\textbf {\bibinfo {volume} {60}},\ \bibinfo {pages} {4170}
  (\bibinfo {year} {1999})}\BibitemShut {NoStop}%
\bibitem [{\citenamefont {Sieling}\ \emph {et~al.}(2000)\citenamefont
  {Sieling}, \citenamefont {L\"ow}, \citenamefont {Wolf}, \citenamefont
  {Schmidt}, \citenamefont {Zvyagin},\ and\ \citenamefont {L\"uthi}}]{Sieling}%
  \BibitemOpen
  \bibfield  {author} {\bibinfo {author} {\bibfnamefont {M.}~\bibnamefont
  {Sieling}}, \bibinfo {author} {\bibfnamefont {U.}~\bibnamefont {L\"ow}},
  \bibinfo {author} {\bibfnamefont {B.}~\bibnamefont {Wolf}}, \bibinfo {author}
  {\bibfnamefont {S.}~\bibnamefont {Schmidt}}, \bibinfo {author} {\bibfnamefont
  {S.}~\bibnamefont {Zvyagin}}, \ and\ \bibinfo {author} {\bibfnamefont
  {B.}~\bibnamefont {L\"uthi}},\ }\href {\doibase 10.1103/PhysRevB.61.88}
  {\bibfield  {journal} {\bibinfo  {journal} {Phys. Rev. B}\ }\textbf {\bibinfo
  {volume} {61}},\ \bibinfo {pages} {88} (\bibinfo {year} {2000})}\BibitemShut
  {NoStop}%
\bibitem [{\citenamefont {Hagiwara}(2002)}]{Hagiwara1}%
  \BibitemOpen
  \bibfield  {author} {\bibinfo {author} {\bibfnamefont {M.}~\bibnamefont
  {Hagiwara}},\ }\href {\doibase 10.1080/00268970110118349} {\bibfield
  {journal} {\bibinfo  {journal} {Molecular Physics}\ }\textbf {\bibinfo
  {volume} {100}},\ \bibinfo {pages} {1489} (\bibinfo {year}
  {2002})}\BibitemShut {NoStop}%
\bibitem [{\citenamefont {\v{C}i\v{z}m\'{a}r}\ \emph
  {et~al.}(2008)\citenamefont {\v{C}i\v{z}m\'{a}r}, \citenamefont {Ozerov},
  \citenamefont {Ignatchik}, \citenamefont {Papageorgiou}, \citenamefont
  {Wosnitza}, \citenamefont {Zvyagin}, \citenamefont {Krzystek}, \citenamefont
  {Zhou}, \citenamefont {Landee}, \citenamefont {Landry}, \citenamefont
  {Turnbull},\ and\ \citenamefont {Wikaira}}]{Cizmar}%
  \BibitemOpen
  \bibfield  {author} {\bibinfo {author} {\bibfnamefont {E.}~\bibnamefont
  {\v{C}i\v{z}m\'{a}r}}, \bibinfo {author} {\bibfnamefont {M.}~\bibnamefont
  {Ozerov}}, \bibinfo {author} {\bibfnamefont {O.}~\bibnamefont {Ignatchik}},
  \bibinfo {author} {\bibfnamefont {T.~P.}\ \bibnamefont {Papageorgiou}},
  \bibinfo {author} {\bibfnamefont {J.}~\bibnamefont {Wosnitza}}, \bibinfo
  {author} {\bibfnamefont {S.~A.}\ \bibnamefont {Zvyagin}}, \bibinfo {author}
  {\bibfnamefont {J.}~\bibnamefont {Krzystek}}, \bibinfo {author}
  {\bibfnamefont {Z.}~\bibnamefont {Zhou}}, \bibinfo {author} {\bibfnamefont
  {C.~P.}\ \bibnamefont {Landee}}, \bibinfo {author} {\bibfnamefont {B.~R.}\
  \bibnamefont {Landry}}, \bibinfo {author} {\bibfnamefont {M.~M.}\
  \bibnamefont {Turnbull}}, \ and\ \bibinfo {author} {\bibfnamefont {J.~L.}\
  \bibnamefont {Wikaira}},\ }\href
  {http://stacks.iop.org/1367-2630/10/i=3/a=033008} {\bibfield  {journal}
  {\bibinfo  {journal} {New Journal of Physics}\ }\textbf {\bibinfo {volume}
  {10}},\ \bibinfo {pages} {033008} (\bibinfo {year} {2008})}\BibitemShut
  {NoStop}%
\bibitem [{\citenamefont {Zheludev}\ \emph {et~al.}(1996)\citenamefont
  {Zheludev}, \citenamefont {Nagler}, \citenamefont {Shapiro}, \citenamefont
  {Chou}, \citenamefont {Talham},\ and\ \citenamefont {Meisel}}]{Zheludev}%
  \BibitemOpen
  \bibfield  {author} {\bibinfo {author} {\bibfnamefont {A.}~\bibnamefont
  {Zheludev}}, \bibinfo {author} {\bibfnamefont {S.~E.}\ \bibnamefont
  {Nagler}}, \bibinfo {author} {\bibfnamefont {S.~M.}\ \bibnamefont {Shapiro}},
  \bibinfo {author} {\bibfnamefont {L.~K.}\ \bibnamefont {Chou}}, \bibinfo
  {author} {\bibfnamefont {D.~R.}\ \bibnamefont {Talham}}, \ and\ \bibinfo
  {author} {\bibfnamefont {M.~W.}\ \bibnamefont {Meisel}},\ }\href {\doibase
  10.1103/PhysRevB.53.15004} {\bibfield  {journal} {\bibinfo  {journal} {Phys.
  Rev. B}\ }\textbf {\bibinfo {volume} {53}},\ \bibinfo {pages} {15004}
  (\bibinfo {year} {1996})}\BibitemShut {NoStop}%
\bibitem [{\citenamefont {Sakai}\ and\ \citenamefont
  {Takahashi}(1990)}]{Sakai}%
  \BibitemOpen
  \bibfield  {author} {\bibinfo {author} {\bibfnamefont {T.}~\bibnamefont
  {Sakai}}\ and\ \bibinfo {author} {\bibfnamefont {M.}~\bibnamefont
  {Takahashi}},\ }\href {\doibase 10.1103/PhysRevB.42.4537} {\bibfield
  {journal} {\bibinfo  {journal} {Phys. Rev. B}\ }\textbf {\bibinfo {volume}
  {42}},\ \bibinfo {pages} {4537} (\bibinfo {year} {1990})}\BibitemShut
  {NoStop}%
\bibitem [{\citenamefont {Tzeng}\ and\ \citenamefont {Yang}(2008)}]{Tzeng}%
  \BibitemOpen
  \bibfield  {author} {\bibinfo {author} {\bibfnamefont {Y.-C.}\ \bibnamefont
  {Tzeng}}\ and\ \bibinfo {author} {\bibfnamefont {M.-F.}\ \bibnamefont
  {Yang}},\ }\href {\doibase 10.1103/PhysRevA.77.012311} {\bibfield  {journal}
  {\bibinfo  {journal} {Phys. Rev. A}\ }\textbf {\bibinfo {volume} {77}},\
  \bibinfo {pages} {012311} (\bibinfo {year} {2008})}\BibitemShut {NoStop}%
\bibitem [{\citenamefont {Albuquerque}\ \emph {et~al.}(2009)\citenamefont
  {Albuquerque}, \citenamefont {Hamer},\ and\ \citenamefont
  {Oitmaa}}]{Albuquerque}%
  \BibitemOpen
  \bibfield  {author} {\bibinfo {author} {\bibfnamefont {A.~F.}\ \bibnamefont
  {Albuquerque}}, \bibinfo {author} {\bibfnamefont {C.~J.}\ \bibnamefont
  {Hamer}}, \ and\ \bibinfo {author} {\bibfnamefont {J.}~\bibnamefont
  {Oitmaa}},\ }\href {\doibase 10.1103/PhysRevB.79.054412} {\bibfield
  {journal} {\bibinfo  {journal} {Phys. Rev. B}\ }\textbf {\bibinfo {volume}
  {79}},\ \bibinfo {pages} {054412} (\bibinfo {year} {2009})}\BibitemShut
  {NoStop}%
\bibitem [{\citenamefont {Pollmann}\ \emph {et~al.}(2010)\citenamefont
  {Pollmann}, \citenamefont {Turner}, \citenamefont {Berg},\ and\ \citenamefont
  {Oshikawa}}]{Pollmann1}%
  \BibitemOpen
  \bibfield  {author} {\bibinfo {author} {\bibfnamefont {F.}~\bibnamefont
  {Pollmann}}, \bibinfo {author} {\bibfnamefont {A.~M.}\ \bibnamefont
  {Turner}}, \bibinfo {author} {\bibfnamefont {E.}~\bibnamefont {Berg}}, \ and\
  \bibinfo {author} {\bibfnamefont {M.}~\bibnamefont {Oshikawa}},\ }\href
  {\doibase 10.1103/PhysRevB.81.064439} {\bibfield  {journal} {\bibinfo
  {journal} {Phys. Rev. B}\ }\textbf {\bibinfo {volume} {81}},\ \bibinfo
  {pages} {064439} (\bibinfo {year} {2010})}\BibitemShut {NoStop}%
\bibitem [{\citenamefont {Hu}\ \emph {et~al.}(2011)\citenamefont {Hu},
  \citenamefont {Normand}, \citenamefont {Wang},\ and\ \citenamefont
  {Yu}}]{Hu}%
  \BibitemOpen
  \bibfield  {author} {\bibinfo {author} {\bibfnamefont {S.}~\bibnamefont
  {Hu}}, \bibinfo {author} {\bibfnamefont {B.}~\bibnamefont {Normand}},
  \bibinfo {author} {\bibfnamefont {X.}~\bibnamefont {Wang}}, \ and\ \bibinfo
  {author} {\bibfnamefont {L.}~\bibnamefont {Yu}},\ }\href {\doibase
  10.1103/PhysRevB.84.220402} {\bibfield  {journal} {\bibinfo  {journal} {Phys.
  Rev. B}\ }\textbf {\bibinfo {volume} {84}},\ \bibinfo {pages} {220402}
  (\bibinfo {year} {2011})}\BibitemShut {NoStop}%
\bibitem [{\citenamefont {Pollmann}\ \emph {et~al.}(2012)\citenamefont
  {Pollmann}, \citenamefont {Berg}, \citenamefont {Turner},\ and\ \citenamefont
  {Oshikawa}}]{Pollmann2}%
  \BibitemOpen
  \bibfield  {author} {\bibinfo {author} {\bibfnamefont {F.}~\bibnamefont
  {Pollmann}}, \bibinfo {author} {\bibfnamefont {E.}~\bibnamefont {Berg}},
  \bibinfo {author} {\bibfnamefont {A.~M.}\ \bibnamefont {Turner}}, \ and\
  \bibinfo {author} {\bibfnamefont {M.}~\bibnamefont {Oshikawa}},\ }\href
  {\doibase 10.1103/PhysRevB.85.075125} {\bibfield  {journal} {\bibinfo
  {journal} {Phys. Rev. B}\ }\textbf {\bibinfo {volume} {85}},\ \bibinfo
  {pages} {075125} (\bibinfo {year} {2012})}\BibitemShut {NoStop}%
\bibitem [{\citenamefont {Li}\ \emph {et~al.}(2013)\citenamefont {Li},
  \citenamefont {Weichselbaum},\ and\ \citenamefont {von Delft}}]{Li}%
  \BibitemOpen
  \bibfield  {author} {\bibinfo {author} {\bibfnamefont {W.}~\bibnamefont
  {Li}}, \bibinfo {author} {\bibfnamefont {A.}~\bibnamefont {Weichselbaum}}, \
  and\ \bibinfo {author} {\bibfnamefont {J.}~\bibnamefont {von Delft}},\ }\href
  {\doibase 10.1103/PhysRevB.88.245121} {\bibfield  {journal} {\bibinfo
  {journal} {Phys. Rev. B}\ }\textbf {\bibinfo {volume} {88}},\ \bibinfo
  {pages} {245121} (\bibinfo {year} {2013})}\BibitemShut {NoStop}%
\bibitem [{\citenamefont {Langari}\ \emph {et~al.}(2013)\citenamefont
  {Langari}, \citenamefont {Pollmann},\ and\ \citenamefont
  {Siahatgar}}]{Langari}%
  \BibitemOpen
  \bibfield  {author} {\bibinfo {author} {\bibfnamefont {A.}~\bibnamefont
  {Langari}}, \bibinfo {author} {\bibfnamefont {F.}~\bibnamefont {Pollmann}}, \
  and\ \bibinfo {author} {\bibfnamefont {M.}~\bibnamefont {Siahatgar}},\ }\href
  {http://stacks.iop.org/0953-8984/25/i=40/a=406002} {\bibfield  {journal}
  {\bibinfo  {journal} {Journal of Physics: Condensed Matter}\ }\textbf
  {\bibinfo {volume} {25}},\ \bibinfo {pages} {406002} (\bibinfo {year}
  {2013})}\BibitemShut {NoStop}%
\bibitem [{\citenamefont {Wierschem}\ and\ \citenamefont
  {Sengupta}(2014)}]{Wierschem}%
  \BibitemOpen
  \bibfield  {author} {\bibinfo {author} {\bibfnamefont {K.}~\bibnamefont
  {Wierschem}}\ and\ \bibinfo {author} {\bibfnamefont {P.}~\bibnamefont
  {Sengupta}},\ }\href {\doibase 10.1103/PhysRevLett.112.247203} {\bibfield
  {journal} {\bibinfo  {journal} {Phys. Rev. Lett.}\ }\textbf {\bibinfo
  {volume} {112}},\ \bibinfo {pages} {247203} (\bibinfo {year}
  {2014})}\BibitemShut {NoStop}%
\bibitem [{\citenamefont {Si}\ and\ \citenamefont {Steglich}(2010)}]{Steglich}%
  \BibitemOpen
  \bibfield  {author} {\bibinfo {author} {\bibfnamefont {Q.}~\bibnamefont
  {Si}}\ and\ \bibinfo {author} {\bibfnamefont {F.}~\bibnamefont {Steglich}},\
  }\href {\doibase 10.1126/science.1191195} {\bibfield  {journal} {\bibinfo
  {journal} {Science}\ }\textbf {\bibinfo {volume} {329}},\ \bibinfo {pages}
  {1161} (\bibinfo {year} {2010})}\BibitemShut {NoStop}%
\bibitem [{\citenamefont {Kinross}\ \emph {et~al.}(2014)\citenamefont
  {Kinross}, \citenamefont {Fu}, \citenamefont {Munsie}, \citenamefont
  {Dabkowska}, \citenamefont {Luke}, \citenamefont {Sachdev},\ and\
  \citenamefont {Imai}}]{Kinross}%
  \BibitemOpen
  \bibfield  {author} {\bibinfo {author} {\bibfnamefont {A.~W.}\ \bibnamefont
  {Kinross}}, \bibinfo {author} {\bibfnamefont {M.}~\bibnamefont {Fu}},
  \bibinfo {author} {\bibfnamefont {T.~J.}\ \bibnamefont {Munsie}}, \bibinfo
  {author} {\bibfnamefont {H.~A.}\ \bibnamefont {Dabkowska}}, \bibinfo {author}
  {\bibfnamefont {G.~M.}\ \bibnamefont {Luke}}, \bibinfo {author}
  {\bibfnamefont {S.}~\bibnamefont {Sachdev}}, \ and\ \bibinfo {author}
  {\bibfnamefont {T.}~\bibnamefont {Imai}},\ }\href {\doibase
  10.1103/PhysRevX.4.031008} {\bibfield  {journal} {\bibinfo  {journal} {Phys.
  Rev. X}\ }\textbf {\bibinfo {volume} {4}},\ \bibinfo {pages} {031008}
  (\bibinfo {year} {2014})}\BibitemShut {NoStop}%
\bibitem [{\citenamefont {Manson}\ \emph {et~al.}(2012)\citenamefont {Manson},
  \citenamefont {Baldwin}, \citenamefont {Scott}, \citenamefont {Bendix},
  \citenamefont {Del~Sesto}, \citenamefont {Goddard}, \citenamefont {Kohama},
  \citenamefont {Tran}, \citenamefont {Ghannadzadeh}, \citenamefont
  {Singleton}, \citenamefont {Lancaster}, \citenamefont {M\"oller},
  \citenamefont {Blundell}, \citenamefont {Pratt}, \citenamefont {Zapf},
  \citenamefont {Kang}, \citenamefont {Lee}, \citenamefont {Whangbo},\ and\
  \citenamefont {Baines}}]{Manson}%
  \BibitemOpen
  \bibfield  {author} {\bibinfo {author} {\bibfnamefont {J.~L.}\ \bibnamefont
  {Manson}}, \bibinfo {author} {\bibfnamefont {A.~G.}\ \bibnamefont {Baldwin}},
  \bibinfo {author} {\bibfnamefont {B.~L.}\ \bibnamefont {Scott}}, \bibinfo
  {author} {\bibfnamefont {J.}~\bibnamefont {Bendix}}, \bibinfo {author}
  {\bibfnamefont {R.~E.}\ \bibnamefont {Del~Sesto}}, \bibinfo {author}
  {\bibfnamefont {P.~A.}\ \bibnamefont {Goddard}}, \bibinfo {author}
  {\bibfnamefont {Y.}~\bibnamefont {Kohama}}, \bibinfo {author} {\bibfnamefont
  {H.~E.}\ \bibnamefont {Tran}}, \bibinfo {author} {\bibfnamefont
  {S.}~\bibnamefont {Ghannadzadeh}}, \bibinfo {author} {\bibfnamefont
  {J.}~\bibnamefont {Singleton}}, \bibinfo {author} {\bibfnamefont
  {T.}~\bibnamefont {Lancaster}}, \bibinfo {author} {\bibfnamefont {J.~S.}\
  \bibnamefont {M\"oller}}, \bibinfo {author} {\bibfnamefont {S.~J.}\
  \bibnamefont {Blundell}}, \bibinfo {author} {\bibfnamefont {F.~L.}\
  \bibnamefont {Pratt}}, \bibinfo {author} {\bibfnamefont {V.~S.}\ \bibnamefont
  {Zapf}}, \bibinfo {author} {\bibfnamefont {J.}~\bibnamefont {Kang}}, \bibinfo
  {author} {\bibfnamefont {C.}~\bibnamefont {Lee}}, \bibinfo {author}
  {\bibfnamefont {M.-H.}\ \bibnamefont {Whangbo}}, \ and\ \bibinfo {author}
  {\bibfnamefont {C.}~\bibnamefont {Baines}},\ }\href {\doibase
  10.1021/ic300111k} {\bibfield  {journal} {\bibinfo  {journal} {Inorganic
  Chemistry}\ }\textbf {\bibinfo {volume} {51}},\ \bibinfo {pages} {7520}
  (\bibinfo {year} {2012})}\BibitemShut {NoStop}%
\bibitem [{\citenamefont {Hagiwara}\ \emph {et~al.}(1990)\citenamefont
  {Hagiwara}, \citenamefont {Katsumata}, \citenamefont {Affleck}, \citenamefont
  {Halperin},\ and\ \citenamefont {Renard}}]{Hagiwara2}%
  \BibitemOpen
  \bibfield  {author} {\bibinfo {author} {\bibfnamefont {M.}~\bibnamefont
  {Hagiwara}}, \bibinfo {author} {\bibfnamefont {K.}~\bibnamefont {Katsumata}},
  \bibinfo {author} {\bibfnamefont {I.}~\bibnamefont {Affleck}}, \bibinfo
  {author} {\bibfnamefont {B.~I.}\ \bibnamefont {Halperin}}, \ and\ \bibinfo
  {author} {\bibfnamefont {J.~P.}\ \bibnamefont {Renard}},\ }\href {\doibase
  10.1103/PhysRevLett.65.3181} {\bibfield  {journal} {\bibinfo  {journal}
  {Phys. Rev. Lett.}\ }\textbf {\bibinfo {volume} {65}},\ \bibinfo {pages}
  {3181} (\bibinfo {year} {1990})}\BibitemShut {NoStop}%
\bibitem [{\citenamefont {Glarum}\ \emph {et~al.}(1991)\citenamefont {Glarum},
  \citenamefont {Geschwind}, \citenamefont {Lee}, \citenamefont {Kaplan},\ and\
  \citenamefont {Michel}}]{Glarum}%
  \BibitemOpen
  \bibfield  {author} {\bibinfo {author} {\bibfnamefont {S.~H.}\ \bibnamefont
  {Glarum}}, \bibinfo {author} {\bibfnamefont {S.}~\bibnamefont {Geschwind}},
  \bibinfo {author} {\bibfnamefont {K.~M.}\ \bibnamefont {Lee}}, \bibinfo
  {author} {\bibfnamefont {M.~L.}\ \bibnamefont {Kaplan}}, \ and\ \bibinfo
  {author} {\bibfnamefont {J.}~\bibnamefont {Michel}},\ }\href {\doibase
  10.1103/PhysRevLett.67.1614} {\bibfield  {journal} {\bibinfo  {journal}
  {Phys. Rev. Lett.}\ }\textbf {\bibinfo {volume} {67}},\ \bibinfo {pages}
  {1614} (\bibinfo {year} {1991})}\BibitemShut {NoStop}%
\bibitem [{\citenamefont {Avenel}\ \emph
  {et~al.}(1992{\natexlab{a}})\citenamefont {Avenel}, \citenamefont {Xu},
  \citenamefont {Xia}, \citenamefont {Xu}, \citenamefont {Andraka},
  \citenamefont {Lang}, \citenamefont {Moyland}, \citenamefont {Ni},
  \citenamefont {Signore}, \citenamefont {van Woerkens}, \citenamefont {Adams},
  \citenamefont {Ihas}, \citenamefont {Meisel}, \citenamefont {Nagler},
  \citenamefont {Sullivan}, \citenamefont {Takano}, \citenamefont {Talham},
  \citenamefont {Goto},\ and\ \citenamefont {Fujiwara}}]{Avenel1}%
  \BibitemOpen
  \bibfield  {author} {\bibinfo {author} {\bibfnamefont {O.}~\bibnamefont
  {Avenel}}, \bibinfo {author} {\bibfnamefont {J.}~\bibnamefont {Xu}}, \bibinfo
  {author} {\bibfnamefont {J.~S.}\ \bibnamefont {Xia}}, \bibinfo {author}
  {\bibfnamefont {M.-F.}\ \bibnamefont {Xu}}, \bibinfo {author} {\bibfnamefont
  {B.}~\bibnamefont {Andraka}}, \bibinfo {author} {\bibfnamefont
  {T.}~\bibnamefont {Lang}}, \bibinfo {author} {\bibfnamefont {P.~L.}\
  \bibnamefont {Moyland}}, \bibinfo {author} {\bibfnamefont {W.}~\bibnamefont
  {Ni}}, \bibinfo {author} {\bibfnamefont {P.~J.~C.}\ \bibnamefont {Signore}},
  \bibinfo {author} {\bibfnamefont {C.~M. C.~M.}\ \bibnamefont {van Woerkens}},
  \bibinfo {author} {\bibfnamefont {E.~D.}\ \bibnamefont {Adams}}, \bibinfo
  {author} {\bibfnamefont {G.~G.}\ \bibnamefont {Ihas}}, \bibinfo {author}
  {\bibfnamefont {M.~W.}\ \bibnamefont {Meisel}}, \bibinfo {author}
  {\bibfnamefont {S.~E.}\ \bibnamefont {Nagler}}, \bibinfo {author}
  {\bibfnamefont {N.~S.}\ \bibnamefont {Sullivan}}, \bibinfo {author}
  {\bibfnamefont {Y.}~\bibnamefont {Takano}}, \bibinfo {author} {\bibfnamefont
  {D.~R.}\ \bibnamefont {Talham}}, \bibinfo {author} {\bibfnamefont
  {T.}~\bibnamefont {Goto}}, \ and\ \bibinfo {author} {\bibfnamefont
  {N.}~\bibnamefont {Fujiwara}},\ }\href {\doibase 10.1103/PhysRevB.46.8655}
  {\bibfield  {journal} {\bibinfo  {journal} {Phys. Rev. B}\ }\textbf {\bibinfo
  {volume} {46}},\ \bibinfo {pages} {8655} (\bibinfo {year}
  {1992}{\natexlab{a}})}\BibitemShut {NoStop}%
\bibitem [{\citenamefont {Avenel}\ \emph
  {et~al.}(1992{\natexlab{b}})\citenamefont {Avenel}, \citenamefont {Xu},
  \citenamefont {Xia}, \citenamefont {Xu}, \citenamefont {Andraka},
  \citenamefont {Lang}, \citenamefont {Moyland}, \citenamefont {Ni},
  \citenamefont {Signore}, \citenamefont {Woerkens}, \citenamefont {Adams},
  \citenamefont {Ihas}, \citenamefont {Meisel}, \citenamefont {Nagler},
  \citenamefont {Sullivan}, \citenamefont {Takano}, \citenamefont {Talham},
  \citenamefont {Goto},\ and\ \citenamefont {Fujiwara}}]{Avenel2}%
  \BibitemOpen
  \bibfield  {author} {\bibinfo {author} {\bibfnamefont {O.}~\bibnamefont
  {Avenel}}, \bibinfo {author} {\bibfnamefont {J.}~\bibnamefont {Xu}}, \bibinfo
  {author} {\bibfnamefont {J.}~\bibnamefont {Xia}}, \bibinfo {author}
  {\bibfnamefont {M.-F.}\ \bibnamefont {Xu}}, \bibinfo {author} {\bibfnamefont
  {B.}~\bibnamefont {Andraka}}, \bibinfo {author} {\bibfnamefont
  {T.}~\bibnamefont {Lang}}, \bibinfo {author} {\bibfnamefont {P.}~\bibnamefont
  {Moyland}}, \bibinfo {author} {\bibfnamefont {W.}~\bibnamefont {Ni}},
  \bibinfo {author} {\bibfnamefont {P.}~\bibnamefont {Signore}}, \bibinfo
  {author} {\bibfnamefont {C.}~\bibnamefont {Woerkens}}, \bibinfo {author}
  {\bibfnamefont {E.}~\bibnamefont {Adams}}, \bibinfo {author} {\bibfnamefont
  {G.}~\bibnamefont {Ihas}}, \bibinfo {author} {\bibfnamefont {M.}~\bibnamefont
  {Meisel}}, \bibinfo {author} {\bibfnamefont {S.}~\bibnamefont {Nagler}},
  \bibinfo {author} {\bibfnamefont {N.}~\bibnamefont {Sullivan}}, \bibinfo
  {author} {\bibfnamefont {Y.}~\bibnamefont {Takano}}, \bibinfo {author}
  {\bibfnamefont {D.}~\bibnamefont {Talham}}, \bibinfo {author} {\bibfnamefont
  {T.}~\bibnamefont {Goto}}, \ and\ \bibinfo {author} {\bibfnamefont
  {N.}~\bibnamefont {Fujiwara}},\ }\href {\doibase 10.1007/BF00694084}
  {\bibfield  {journal} {\bibinfo  {journal} {Journal of Low Temperature
  Physics}\ }\textbf {\bibinfo {volume} {89}},\ \bibinfo {pages} {547}
  (\bibinfo {year} {1992}{\natexlab{b}})}\BibitemShut {NoStop}%
\bibitem [{\citenamefont {Batista}\ \emph {et~al.}(1998)\citenamefont
  {Batista}, \citenamefont {Hallberg},\ and\ \citenamefont {Aligia}}]{Batista}%
  \BibitemOpen
  \bibfield  {author} {\bibinfo {author} {\bibfnamefont {C.~D.}\ \bibnamefont
  {Batista}}, \bibinfo {author} {\bibfnamefont {K.}~\bibnamefont {Hallberg}}, \
  and\ \bibinfo {author} {\bibfnamefont {A.~A.}\ \bibnamefont {Aligia}},\
  }\href {\doibase 10.1103/PhysRevB.58.9248} {\bibfield  {journal} {\bibinfo
  {journal} {Phys. Rev. B}\ }\textbf {\bibinfo {volume} {58}},\ \bibinfo
  {pages} {9248} (\bibinfo {year} {1998})}\BibitemShut {NoStop}%
\bibitem [{\citenamefont {Granroth}\ \emph {et~al.}(1998)\citenamefont
  {Granroth}, \citenamefont {Maegawa}, \citenamefont {Meisel}, \citenamefont
  {Krzystek}, \citenamefont {Brunel}, \citenamefont {Bell}, \citenamefont
  {Adair}, \citenamefont {Ward}, \citenamefont {Fanucci}, \citenamefont
  {Chou},\ and\ \citenamefont {Talham}}]{Granroth}%
  \BibitemOpen
  \bibfield  {author} {\bibinfo {author} {\bibfnamefont {G.~E.}\ \bibnamefont
  {Granroth}}, \bibinfo {author} {\bibfnamefont {S.}~\bibnamefont {Maegawa}},
  \bibinfo {author} {\bibfnamefont {M.~W.}\ \bibnamefont {Meisel}}, \bibinfo
  {author} {\bibfnamefont {J.}~\bibnamefont {Krzystek}}, \bibinfo {author}
  {\bibfnamefont {L.-C.}\ \bibnamefont {Brunel}}, \bibinfo {author}
  {\bibfnamefont {N.~S.}\ \bibnamefont {Bell}}, \bibinfo {author}
  {\bibfnamefont {J.~H.}\ \bibnamefont {Adair}}, \bibinfo {author}
  {\bibfnamefont {B.~H.}\ \bibnamefont {Ward}}, \bibinfo {author}
  {\bibfnamefont {G.~E.}\ \bibnamefont {Fanucci}}, \bibinfo {author}
  {\bibfnamefont {L.-K.}\ \bibnamefont {Chou}}, \ and\ \bibinfo {author}
  {\bibfnamefont {D.~R.}\ \bibnamefont {Talham}},\ }\href {\doibase
  10.1103/PhysRevB.58.9312} {\bibfield  {journal} {\bibinfo  {journal} {Phys.
  Rev. B}\ }\textbf {\bibinfo {volume} {58}},\ \bibinfo {pages} {9312}
  (\bibinfo {year} {1998})}\BibitemShut {NoStop}%
\bibitem [{\citenamefont {Chiba}\ \emph {et~al.}(1991)\citenamefont {Chiba},
  \citenamefont {Ajiro}, \citenamefont {Kikuchi}, \citenamefont {Kubo},\ and\
  \citenamefont {Morimoto}}]{Chiba}%
  \BibitemOpen
  \bibfield  {author} {\bibinfo {author} {\bibfnamefont {M.}~\bibnamefont
  {Chiba}}, \bibinfo {author} {\bibfnamefont {Y.}~\bibnamefont {Ajiro}},
  \bibinfo {author} {\bibfnamefont {H.}~\bibnamefont {Kikuchi}}, \bibinfo
  {author} {\bibfnamefont {T.}~\bibnamefont {Kubo}}, \ and\ \bibinfo {author}
  {\bibfnamefont {T.}~\bibnamefont {Morimoto}},\ }\href {\doibase
  10.1103/PhysRevB.44.2838} {\bibfield  {journal} {\bibinfo  {journal} {Phys.
  Rev. B}\ }\textbf {\bibinfo {volume} {44}},\ \bibinfo {pages} {2838}
  (\bibinfo {year} {1991})}\BibitemShut {NoStop}%
\bibitem [{\citenamefont {Huang}\ and\ \citenamefont
  {Affleck}(2004)}]{Affleck-staggered}%
  \BibitemOpen
  \bibfield  {author} {\bibinfo {author} {\bibfnamefont {H.}~\bibnamefont
  {Huang}}\ and\ \bibinfo {author} {\bibfnamefont {I.}~\bibnamefont
  {Affleck}},\ }\href {\doibase 10.1103/PhysRevB.69.184414} {\bibfield
  {journal} {\bibinfo  {journal} {Phys. Rev. B}\ }\textbf {\bibinfo {volume}
  {69}},\ \bibinfo {pages} {184414} (\bibinfo {year} {2004})}\BibitemShut
  {NoStop}%
\bibitem [{\citenamefont {Hammel}\ \emph {et~al.}(1983)\citenamefont {Hammel},
  \citenamefont {Roukes}, \citenamefont {Hu}, \citenamefont {Gramila},
  \citenamefont {Mamiya},\ and\ \citenamefont {Richardson}}]{Hammel}%
  \BibitemOpen
  \bibfield  {author} {\bibinfo {author} {\bibfnamefont {P.~C.}\ \bibnamefont
  {Hammel}}, \bibinfo {author} {\bibfnamefont {M.~L.}\ \bibnamefont {Roukes}},
  \bibinfo {author} {\bibfnamefont {Y.}~\bibnamefont {Hu}}, \bibinfo {author}
  {\bibfnamefont {T.~J.}\ \bibnamefont {Gramila}}, \bibinfo {author}
  {\bibfnamefont {T.}~\bibnamefont {Mamiya}}, \ and\ \bibinfo {author}
  {\bibfnamefont {R.~C.}\ \bibnamefont {Richardson}},\ }\href {\doibase
  10.1103/PhysRevLett.51.2124} {\bibfield  {journal} {\bibinfo  {journal}
  {Phys. Rev. Lett.}\ }\textbf {\bibinfo {volume} {51}},\ \bibinfo {pages}
  {2124} (\bibinfo {year} {1983})}\BibitemShut {NoStop}%
\bibitem [{\citenamefont {Schuhl}\ \emph {et~al.}(1987)\citenamefont {Schuhl},
  \citenamefont {Maegawa}, \citenamefont {Meisel},\ and\ \citenamefont
  {Chapellier}}]{Schuhl}%
  \BibitemOpen
  \bibfield  {author} {\bibinfo {author} {\bibfnamefont {A.}~\bibnamefont
  {Schuhl}}, \bibinfo {author} {\bibfnamefont {S.}~\bibnamefont {Maegawa}},
  \bibinfo {author} {\bibfnamefont {M.~W.}\ \bibnamefont {Meisel}}, \ and\
  \bibinfo {author} {\bibfnamefont {M.}~\bibnamefont {Chapellier}},\ }\href
  {\doibase 10.1103/PhysRevB.36.6811} {\bibfield  {journal} {\bibinfo
  {journal} {Phys. Rev. B}\ }\textbf {\bibinfo {volume} {36}},\ \bibinfo
  {pages} {6811} (\bibinfo {year} {1987})}\BibitemShut {NoStop}%
\bibitem [{\citenamefont {Orend\'a\v{c}ov\'a}\ \emph
  {et~al.}(2009)\citenamefont {Orend\'a\v{c}ov\'a}, \citenamefont
  {\v{C}i\v{z}m\'ar}, \citenamefont {Sedl\'akov\'a}, \citenamefont {Hanko},
  \citenamefont {Kaj\v{n}akov\'a}, \citenamefont {Orend\'a\v{c}}, \citenamefont
  {Feher}, \citenamefont {Xia}, \citenamefont {Yin}, \citenamefont
  {Pajerowski}, \citenamefont {Meisel}, \citenamefont {Zele\v{n}\'ak},
  \citenamefont {Zvyagin},\ and\ \citenamefont {Wosnitza}}]{Orendacova}%
  \BibitemOpen
  \bibfield  {author} {\bibinfo {author} {\bibfnamefont {A.}~\bibnamefont
  {Orend\'a\v{c}ov\'a}}, \bibinfo {author} {\bibfnamefont {E.}~\bibnamefont
  {\v{C}i\v{z}m\'ar}}, \bibinfo {author} {\bibfnamefont {L.}~\bibnamefont
  {Sedl\'akov\'a}}, \bibinfo {author} {\bibfnamefont {J.}~\bibnamefont
  {Hanko}}, \bibinfo {author} {\bibfnamefont {M.}~\bibnamefont
  {Kaj\v{n}akov\'a}}, \bibinfo {author} {\bibfnamefont {M.}~\bibnamefont
  {Orend\'a\v{c}}}, \bibinfo {author} {\bibfnamefont {A.}~\bibnamefont
  {Feher}}, \bibinfo {author} {\bibfnamefont {J.~S.}\ \bibnamefont {Xia}},
  \bibinfo {author} {\bibfnamefont {L.}~\bibnamefont {Yin}}, \bibinfo {author}
  {\bibfnamefont {D.~M.}\ \bibnamefont {Pajerowski}}, \bibinfo {author}
  {\bibfnamefont {M.~W.}\ \bibnamefont {Meisel}}, \bibinfo {author}
  {\bibfnamefont {V.}~\bibnamefont {Zele\v{n}\'ak}}, \bibinfo {author}
  {\bibfnamefont {S.}~\bibnamefont {Zvyagin}}, \ and\ \bibinfo {author}
  {\bibfnamefont {J.}~\bibnamefont {Wosnitza}},\ }\href {\doibase
  10.1103/PhysRevB.80.144418} {\bibfield  {journal} {\bibinfo  {journal} {Phys.
  Rev. B}\ }\textbf {\bibinfo {volume} {80}},\ \bibinfo {pages} {144418}
  (\bibinfo {year} {2009})}\BibitemShut {NoStop}%
\bibitem [{\citenamefont {Vidal}\ \emph {et~al.}(2003)\citenamefont {Vidal},
  \citenamefont {Latorre}, \citenamefont {Rico},\ and\ \citenamefont
  {Kitaev}}]{Vidal}%
  \BibitemOpen
  \bibfield  {author} {\bibinfo {author} {\bibfnamefont {G.}~\bibnamefont
  {Vidal}}, \bibinfo {author} {\bibfnamefont {J.~I.}\ \bibnamefont {Latorre}},
  \bibinfo {author} {\bibfnamefont {E.}~\bibnamefont {Rico}}, \ and\ \bibinfo
  {author} {\bibfnamefont {A.}~\bibnamefont {Kitaev}},\ }\href {\doibase
  10.1103/PhysRevLett.90.227902} {\bibfield  {journal} {\bibinfo  {journal}
  {Phys. Rev. Lett.}\ }\textbf {\bibinfo {volume} {90}},\ \bibinfo {pages}
  {227902} (\bibinfo {year} {2003})}\BibitemShut {NoStop}%
\bibitem [{\citenamefont {Zhu}\ \emph {et~al.}(2003)\citenamefont {Zhu},
  \citenamefont {Garst}, \citenamefont {Rosch},\ and\ \citenamefont
  {Si}}]{Si2}%
  \BibitemOpen
  \bibfield  {author} {\bibinfo {author} {\bibfnamefont {L.}~\bibnamefont
  {Zhu}}, \bibinfo {author} {\bibfnamefont {M.}~\bibnamefont {Garst}}, \bibinfo
  {author} {\bibfnamefont {A.}~\bibnamefont {Rosch}}, \ and\ \bibinfo {author}
  {\bibfnamefont {Q.}~\bibnamefont {Si}},\ }\href {\doibase
  10.1103/PhysRevLett.91.066404} {\bibfield  {journal} {\bibinfo  {journal}
  {Phys. Rev. Lett.}\ }\textbf {\bibinfo {volume} {91}},\ \bibinfo {pages}
  {066404} (\bibinfo {year} {2003})}\BibitemShut {NoStop}%
\bibitem [{\citenamefont {Hassan}\ \emph {et~al.}(2000)\citenamefont {Hassan},
  \citenamefont {Pardi}, \citenamefont {Krzystek}, \citenamefont {Sienkiewicz},
  \citenamefont {Goy}, \citenamefont {Rohrer},\ and\ \citenamefont
  {Brunel}}]{Hassan}%
  \BibitemOpen
  \bibfield  {author} {\bibinfo {author} {\bibfnamefont {A.}~\bibnamefont
  {Hassan}}, \bibinfo {author} {\bibfnamefont {L.}~\bibnamefont {Pardi}},
  \bibinfo {author} {\bibfnamefont {J.}~\bibnamefont {Krzystek}}, \bibinfo
  {author} {\bibfnamefont {A.}~\bibnamefont {Sienkiewicz}}, \bibinfo {author}
  {\bibfnamefont {P.}~\bibnamefont {Goy}}, \bibinfo {author} {\bibfnamefont
  {M.}~\bibnamefont {Rohrer}}, \ and\ \bibinfo {author} {\bibfnamefont {L.-C.}\
  \bibnamefont {Brunel}},\ }\href {\doibase
  http://dx.doi.org/10.1006/jmre.1999.1952} {\bibfield  {journal} {\bibinfo
  {journal} {Journal of Magnetic Resonance}\ }\textbf {\bibinfo {volume}
  {142}},\ \bibinfo {pages} {300 } (\bibinfo {year} {2000})}\BibitemShut
  {NoStop}%
\bibitem [{\citenamefont {Hagiwara}\ \emph {et~al.}(1996)\citenamefont
  {Hagiwara}, \citenamefont {Katsumata}, \citenamefont {Yamada},\ and\
  \citenamefont {Suzuki}}]{HagiwaraMnF2}%
  \BibitemOpen
  \bibfield  {author} {\bibinfo {author} {\bibfnamefont {M.}~\bibnamefont
  {Hagiwara}}, \bibinfo {author} {\bibfnamefont {K.}~\bibnamefont {Katsumata}},
  \bibinfo {author} {\bibfnamefont {I.}~\bibnamefont {Yamada}}, \ and\ \bibinfo
  {author} {\bibfnamefont {H.}~\bibnamefont {Suzuki}},\ }\href
  {http://stacks.iop.org/0953-8984/8/i=39/a=011} {\bibfield  {journal}
  {\bibinfo  {journal} {Journal of Physics: Condensed Matter}\ }\textbf
  {\bibinfo {volume} {8}},\ \bibinfo {pages} {7349} (\bibinfo {year}
  {1996})}\BibitemShut {NoStop}%
\bibitem [{\citenamefont {Fanucci}\ \emph {et~al.}(1998)\citenamefont
  {Fanucci}, \citenamefont {Krzystek}, \citenamefont {Meisel}, \citenamefont
  {Brunel},\ and\ \citenamefont {Talham}}]{Fanucci}%
  \BibitemOpen
  \bibfield  {author} {\bibinfo {author} {\bibfnamefont {G.~E.}\ \bibnamefont
  {Fanucci}}, \bibinfo {author} {\bibfnamefont {J.}~\bibnamefont {Krzystek}},
  \bibinfo {author} {\bibfnamefont {M.~W.}\ \bibnamefont {Meisel}}, \bibinfo
  {author} {\bibfnamefont {L.-C.}\ \bibnamefont {Brunel}}, \ and\ \bibinfo
  {author} {\bibfnamefont {D.~R.}\ \bibnamefont {Talham}},\ }\href {\doibase
  10.1021/ja974247g} {\bibfield  {journal} {\bibinfo  {journal} {Journal of the
  American Chemical Society}\ }\textbf {\bibinfo {volume} {120}},\ \bibinfo
  {pages} {5469} (\bibinfo {year} {1998})}\BibitemShut {NoStop}%
\bibitem [{\citenamefont {Furuya}\ \emph {et~al.}(2011)\citenamefont {Furuya},
  \citenamefont {Oshikawa},\ and\ \citenamefont
  {Affleck}}]{PhysRevB.83.224417}%
  \BibitemOpen
  \bibfield  {author} {\bibinfo {author} {\bibfnamefont {S.~C.}\ \bibnamefont
  {Furuya}}, \bibinfo {author} {\bibfnamefont {M.}~\bibnamefont {Oshikawa}}, \
  and\ \bibinfo {author} {\bibfnamefont {I.}~\bibnamefont {Affleck}},\ }\href
  {\doibase 10.1103/PhysRevB.83.224417} {\bibfield  {journal} {\bibinfo
  {journal} {Phys. Rev. B}\ }\textbf {\bibinfo {volume} {83}},\ \bibinfo
  {pages} {224417} (\bibinfo {year} {2011})}\BibitemShut {NoStop}%
\bibitem [{\citenamefont {Zaliznyak}\ \emph {et~al.}(1998)\citenamefont
  {Zaliznyak}, \citenamefont {Dender}, \citenamefont {Broholm},\ and\
  \citenamefont {Reich}}]{Broholm}%
  \BibitemOpen
  \bibfield  {author} {\bibinfo {author} {\bibfnamefont {I.~A.}\ \bibnamefont
  {Zaliznyak}}, \bibinfo {author} {\bibfnamefont {D.~C.}\ \bibnamefont
  {Dender}}, \bibinfo {author} {\bibfnamefont {C.}~\bibnamefont {Broholm}}, \
  and\ \bibinfo {author} {\bibfnamefont {D.~H.}\ \bibnamefont {Reich}},\ }\href
  {\doibase 10.1103/PhysRevB.57.5200} {\bibfield  {journal} {\bibinfo
  {journal} {Phys. Rev. B}\ }\textbf {\bibinfo {volume} {57}},\ \bibinfo
  {pages} {5200} (\bibinfo {year} {1998})}\BibitemShut {NoStop}%
\bibitem [{\citenamefont {Oosawa}\ \emph {et~al.}(2004)\citenamefont {Oosawa},
  \citenamefont {Kakurai}, \citenamefont {Osakabe}, \citenamefont {Nakamura},
  \citenamefont {Takeda},\ and\ \citenamefont {Tanaka}}]{Oosawa}%
  \BibitemOpen
  \bibfield  {author} {\bibinfo {author} {\bibfnamefont {A.}~\bibnamefont
  {Oosawa}}, \bibinfo {author} {\bibfnamefont {K.}~\bibnamefont {Kakurai}},
  \bibinfo {author} {\bibfnamefont {T.}~\bibnamefont {Osakabe}}, \bibinfo
  {author} {\bibfnamefont {M.}~\bibnamefont {Nakamura}}, \bibinfo {author}
  {\bibfnamefont {M.}~\bibnamefont {Takeda}}, \ and\ \bibinfo {author}
  {\bibfnamefont {H.}~\bibnamefont {Tanaka}},\ }\href {\doibase
  10.1143/JPSJ.73.1446} {\bibfield  {journal} {\bibinfo  {journal} {Journal of
  the Physical Society of Japan}\ }\textbf {\bibinfo {volume} {73}},\ \bibinfo
  {pages} {1446} (\bibinfo {year} {2004})}\BibitemShut {NoStop}%
\bibitem [{\citenamefont {R\"uegg}\ \emph {et~al.}(2004)\citenamefont
  {R\"uegg}, \citenamefont {Furrer}, \citenamefont {Sheptyakov}, \citenamefont
  {Str\"assle}, \citenamefont {Kr\"amer}, \citenamefont {G\"udel},\ and\
  \citenamefont {M\'el\'esi}}]{Ruegg}%
  \BibitemOpen
  \bibfield  {author} {\bibinfo {author} {\bibfnamefont {C.}~\bibnamefont
  {R\"uegg}}, \bibinfo {author} {\bibfnamefont {A.}~\bibnamefont {Furrer}},
  \bibinfo {author} {\bibfnamefont {D.}~\bibnamefont {Sheptyakov}}, \bibinfo
  {author} {\bibfnamefont {T.}~\bibnamefont {Str\"assle}}, \bibinfo {author}
  {\bibfnamefont {K.~W.}\ \bibnamefont {Kr\"amer}}, \bibinfo {author}
  {\bibfnamefont {H.-U.}\ \bibnamefont {G\"udel}}, \ and\ \bibinfo {author}
  {\bibfnamefont {L.}~\bibnamefont {M\'el\'esi}},\ }\href {\doibase
  10.1103/PhysRevLett.93.257201} {\bibfield  {journal} {\bibinfo  {journal}
  {Phys. Rev. Lett.}\ }\textbf {\bibinfo {volume} {93}},\ \bibinfo {pages}
  {257201} (\bibinfo {year} {2004})}\BibitemShut {NoStop}%
\bibitem [{\citenamefont {Hong}\ \emph {et~al.}(2008)\citenamefont {Hong},
  \citenamefont {Garlea}, \citenamefont {Zheludev}, \citenamefont
  {Fernandez-Baca}, \citenamefont {Manaka}, \citenamefont {Chang},
  \citenamefont {Leao},\ and\ \citenamefont {Poulton}}]{Tao}%
  \BibitemOpen
  \bibfield  {author} {\bibinfo {author} {\bibfnamefont {T.}~\bibnamefont
  {Hong}}, \bibinfo {author} {\bibfnamefont {V.~O.}\ \bibnamefont {Garlea}},
  \bibinfo {author} {\bibfnamefont {A.}~\bibnamefont {Zheludev}}, \bibinfo
  {author} {\bibfnamefont {J.~A.}\ \bibnamefont {Fernandez-Baca}}, \bibinfo
  {author} {\bibfnamefont {H.}~\bibnamefont {Manaka}}, \bibinfo {author}
  {\bibfnamefont {S.}~\bibnamefont {Chang}}, \bibinfo {author} {\bibfnamefont
  {J.~B.}\ \bibnamefont {Leao}}, \ and\ \bibinfo {author} {\bibfnamefont
  {S.~J.}\ \bibnamefont {Poulton}},\ }\href {\doibase
  10.1103/PhysRevB.78.224409} {\bibfield  {journal} {\bibinfo  {journal} {Phys.
  Rev. B}\ }\textbf {\bibinfo {volume} {78}},\ \bibinfo {pages} {224409}
  (\bibinfo {year} {2008})}\BibitemShut {NoStop}%
\bibitem [{\citenamefont {Thede}\ \emph {et~al.}(2014)\citenamefont {Thede},
  \citenamefont {Mannig}, \citenamefont {M\aa{}nsson}, \citenamefont
  {H\"uvonen}, \citenamefont {Khasanov}, \citenamefont {Morenzoni},\ and\
  \citenamefont {Zheludev}}]{Zheludev2}%
  \BibitemOpen
  \bibfield  {author} {\bibinfo {author} {\bibfnamefont {M.}~\bibnamefont
  {Thede}}, \bibinfo {author} {\bibfnamefont {A.}~\bibnamefont {Mannig}},
  \bibinfo {author} {\bibfnamefont {M.}~\bibnamefont {M\aa{}nsson}}, \bibinfo
  {author} {\bibfnamefont {D.}~\bibnamefont {H\"uvonen}}, \bibinfo {author}
  {\bibfnamefont {R.}~\bibnamefont {Khasanov}}, \bibinfo {author}
  {\bibfnamefont {E.}~\bibnamefont {Morenzoni}}, \ and\ \bibinfo {author}
  {\bibfnamefont {A.}~\bibnamefont {Zheludev}},\ }\href {\doibase
  10.1103/PhysRevLett.112.087204} {\bibfield  {journal} {\bibinfo  {journal}
  {Phys. Rev. Lett.}\ }\textbf {\bibinfo {volume} {112}},\ \bibinfo {pages}
  {087204} (\bibinfo {year} {2014})}\BibitemShut {NoStop}%
\bibitem [{\citenamefont {Tsujii}\ \emph {et~al.}(2005)\citenamefont {Tsujii},
  \citenamefont {Honda}, \citenamefont {Andraka}, \citenamefont {Katsumata},\
  and\ \citenamefont {Takano}}]{Takano}%
  \BibitemOpen
  \bibfield  {author} {\bibinfo {author} {\bibfnamefont {H.}~\bibnamefont
  {Tsujii}}, \bibinfo {author} {\bibfnamefont {Z.}~\bibnamefont {Honda}},
  \bibinfo {author} {\bibfnamefont {B.}~\bibnamefont {Andraka}}, \bibinfo
  {author} {\bibfnamefont {K.}~\bibnamefont {Katsumata}}, \ and\ \bibinfo
  {author} {\bibfnamefont {Y.}~\bibnamefont {Takano}},\ }\href {\doibase
  10.1103/PhysRevB.71.014426} {\bibfield  {journal} {\bibinfo  {journal} {Phys.
  Rev. B}\ }\textbf {\bibinfo {volume} {71}},\ \bibinfo {pages} {014426}
  (\bibinfo {year} {2005})}\BibitemShut {NoStop}%
\bibitem [{\citenamefont {Goddard}\ \emph {et~al.}(2008)\citenamefont
  {Goddard}, \citenamefont {Singleton}, \citenamefont {Maitland}, \citenamefont
  {Blundell}, \citenamefont {Lancaster}, \citenamefont {Baker}, \citenamefont
  {McDonald}, \citenamefont {Cox}, \citenamefont {Sengupta}, \citenamefont
  {Manson}, \citenamefont {Funk},\ and\ \citenamefont {Schlueter}}]{Goddard}%
  \BibitemOpen
  \bibfield  {author} {\bibinfo {author} {\bibfnamefont {P.~A.}\ \bibnamefont
  {Goddard}}, \bibinfo {author} {\bibfnamefont {J.}~\bibnamefont {Singleton}},
  \bibinfo {author} {\bibfnamefont {C.}~\bibnamefont {Maitland}}, \bibinfo
  {author} {\bibfnamefont {S.~J.}\ \bibnamefont {Blundell}}, \bibinfo {author}
  {\bibfnamefont {T.}~\bibnamefont {Lancaster}}, \bibinfo {author}
  {\bibfnamefont {P.~J.}\ \bibnamefont {Baker}}, \bibinfo {author}
  {\bibfnamefont {R.~D.}\ \bibnamefont {McDonald}}, \bibinfo {author}
  {\bibfnamefont {S.}~\bibnamefont {Cox}}, \bibinfo {author} {\bibfnamefont
  {P.}~\bibnamefont {Sengupta}}, \bibinfo {author} {\bibfnamefont {J.~L.}\
  \bibnamefont {Manson}}, \bibinfo {author} {\bibfnamefont {K.~A.}\
  \bibnamefont {Funk}}, \ and\ \bibinfo {author} {\bibfnamefont {J.~A.}\
  \bibnamefont {Schlueter}},\ }\href {\doibase 10.1103/PhysRevB.78.052408}
  {\bibfield  {journal} {\bibinfo  {journal} {Phys. Rev. B}\ }\textbf {\bibinfo
  {volume} {78}},\ \bibinfo {pages} {052408} (\bibinfo {year}
  {2008})}\BibitemShut {NoStop}%
\bibitem [{\citenamefont {Pardi}\ \emph {et~al.}(2000)\citenamefont {Pardi},
  \citenamefont {Krzystek}, \citenamefont {Telser},\ and\ \citenamefont
  {Brunel}}]{Pardi}%
  \BibitemOpen
  \bibfield  {author} {\bibinfo {author} {\bibfnamefont {L.~A.}\ \bibnamefont
  {Pardi}}, \bibinfo {author} {\bibfnamefont {J.}~\bibnamefont {Krzystek}},
  \bibinfo {author} {\bibfnamefont {J.}~\bibnamefont {Telser}}, \ and\ \bibinfo
  {author} {\bibfnamefont {L.-C.}\ \bibnamefont {Brunel}},\ }\href {\doibase
  http://dx.doi.org/10.1006/jmre.2000.2175} {\bibfield  {journal} {\bibinfo
  {journal} {Journal of Magnetic Resonance}\ }\textbf {\bibinfo {volume}
  {146}},\ \bibinfo {pages} {375 } (\bibinfo {year} {2000})}\BibitemShut
  {NoStop}%
\end{thebibliography}%

\end{document}